\documentclass[aps,prl,reprint,groupedaddress]{revtex4-1}

\usepackage{amsmath,amssymb}
\usepackage{amsfonts}
\usepackage[squaren,cdot]{SIunits}
\usepackage{graphicx}
\usepackage{color}

\newcommand{\Ca}{C\!a} % pour tracer Ca sans espace entre C et a
\begin{document}

\title{Extension of a suspended soap film: a two-step process}

\author{Jacopo Seiwert}
\author{Martin Monloubou}
\author{Benjamin Dollet}
\author{Isabelle Cantat}
\affiliation{Institut de Physique de Rennes, UMR 6251 CNRS / Universit\'e de Rennes 1, Rennes, France}

\date{\today}

\begin{abstract}
Liquid foams are widely used in industry for their high effective viscosity, whose local origin is still unclear. 
This paper presents new results on the extension of a suspended soap film, in a configuration mimicking the elementary deformation occurring during foam shearing. We evidence a surprising two-step evolution: the film first extends homogeneously, then its extension stops and a new thicker film  is extracted from the meniscus. The second step is independent of the nature of the surfactant solution, 
whereas the initial extension is only observed for surfactant solutions with negligible dilatational moduli.
We predict this complex behavior using a model based on Frankel's theory, and on interface rigidification induced by confinement.
\end{abstract}


\maketitle
Liquid foams are very dissipative materials, with an exceptionally high ratio between their effective viscosity and their density. For this reason, they are widely used for blast wave mitigation \cite{britan13,delprete13}, or as drilling fluid in the oil industry \cite{stevenson}, among other examples. However, the local processes leading to this high energy dissipation have not been entirely elucidated yet \cite{cohenaddad13}. 
The dissipation rate is strongly enhanced by the confinement of the viscous liquid phase in thin films and menisci and it is extremely sensitive to the interfacial properties.
Consequently it depends on the local film thickness and on the interfacial stresses and velocities, acting as boundary conditions for the flows. These quantities are in practice impossible to measure {\it in situ} in a 3D sheared foam, making the prediction of the dissipation rate in such system particularly challenging. 

Local flow models at the bubble scale have been developed, leading to predictions at the foam scale which can be compared to experimental data \cite{kraynik87,schwartz87,buzza95,denkov08}. However, direct measurements of the film thicknesses or deformations at the bubble scale are still sparse \cite{durand06,denkov06,besson07,biance09,dollet10,biance11,cantat12}, although they are necessary to discriminate the very different assumptions made by the models on the local flow.
Recent studies have established the relation between the interfacial rheology of foaming solutions and the bulk rheology of foams \cite{denkov05,marze08, denkov09b}. Interfacial rheology may be characterized by a dilatational modulus $E_d$, which relates relative interface area variations to interfacial stresses, and which can span several orders of magnitudes. Two limiting cases are traditionally considered: (i) ``rigid"  surfactants feature large values of $E_d$ (when compared to the surface tension), and exhibit incompressible interfaces \cite{golemanov08}, while (ii) ``mobile" surfactants exhibit stress-free interfaces.

Physically, interfacial stresses mainly arise because stretching (or compressing) an interface leads to changes in the interfacial concentration of surfactants, hence to surface tension gradients. These gradients relax through the exchange of surfactants between the interface and the bulk, whose typical timescale depends on several processes: adsorption/desorption of surfactant molecules, diffusion, micelles dynamics, etc. $E_d$ is thus not an intrinsic property of the solution, but depends on the time and space scales of the method of measurement. From a practical standpoint, it is still unclear which of these two limiting models (incompressible and stress-free), if any, will apply for a given  experimental situation and a given surfactant solution, because the relevant  value of $E_d$ is in general unknown. 

Film area increase is one of the key processes occurring during foam shearing, along with film formation (after a bubble rearrangement), film shearing and film area decrease.
The linear response of a liquid film under oscillatory extensional strain has been precisely measured \cite{besson07,besson08}, but  large extensional deformations involve entirely different, nonlinear, processes which are the scope of this Letter.
We show that the area increase of a preexisting thin film subjected to a large extension is mainly due to the extraction of a fresh film from the meniscus, and not to an uniform extension of the preexisting film. 
The fresh film thickness quantitatively obeys Frankel's law, which quantifies the extraction of a soap film from a bath assuming incompressible interfaces \cite{mysels}.
This behavior is observed for all tested surfactant solutions, independently of their dilatational moduli.
However, the area of the preexisting film slightly increases during a first step, with a relative area variation that strongly depends on the nature and on the concentration  of the surfactants. We argue that this initial extension modifies the surface tension in the film and thus produces the resistance required to pull the Frankel's film out of the meniscus. This two-step process evidences a transition between a compressible behavior of the interfaces at small strain and an incompressible behavior at larger strain for solutions usually leading to mobile interfaces without measurable resistance to compression/dilatation.

		\begin{figure}[htp!]
	   		\includegraphics[width=8.6cm]{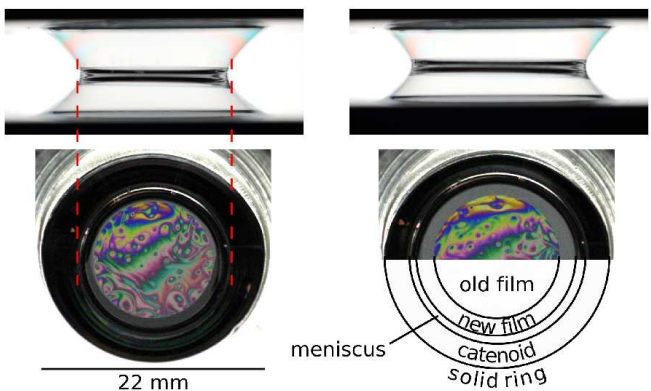}
			\caption{Top: side view of the experiment before (left) and during (right) stage motion. The darker band in the center of the image is the Plateau border. Bottom: corresponding top views, under white illumination (used for illustration: the thickness measurements are made with monochromatic light). It shows the old thin film with heterogeneous bright colors (online), surrounded by the thicker new film appearing uniformly gray. 
			\label{fig:experiment}}
		\end{figure}

The experimental setup, shown on Fig.~\ref{fig:experiment}, features three axisymmetric soap films: a circular horizontal film which we study, and two catenoids used to suspend it between two metallic circular rings (\unit{22}{\milli \meter} in diameter). The radius of the horizontal film increases when the distance between the rings decreases \cite{besson07,geminard04}. These three films meet at a meniscus (Plateau border), as within liquid foams. 

A typical experiment proceeds as follows: the films are created, then let at rest. The pressure in the concave Plateau border is lower than in the flat thin films because of the Laplace pressure, inducing slow film drainage. The horizontal film thickness reaches a quasi-uniform value $h_0$, in the range 0.5-\unit{1.5}{\micro \meter}, after typically 20-\unit{60}{\second}.
The horizontal film radius $R$ is then increased from \unit{7}{\milli \meter} to \unit{10}{\milli \meter} by moving the lower ring at a constant velocity $U_s$ in the range 0.05-\unit{50}{\milli \meter \per \second}.  The lateral film profile is monitored by a first camera  (see Fig.~\ref{fig:experiment}, top).
The interference pattern produced in the horizontal film by a monochromatic lamp  is recorded with a second camera  (Fig.~\ref{fig:experiment}, bottom), giving a time-resolved map of the relative thickness of the film. A local absolute thickness reference is provided by a synchronized reflective interferometer.
Additionally the small thickness fluctuations act as passive tracers \cite{chomaz01} and provide qualitative informations on the velocity field in the film. 

The radius of curvature $r_m$ of the Plateau border was varied between 0.2 and \unit{2.0}{\milli \meter} by adding or withdrawing liquid with a syringe.
 We used three solutions (T3, T13, T25, see Table~\ref{tab:solutions}) with tetradecyltrimethylammonium bromide (TTAB) as surfactant.
TTAB is a soluble surfactant and has a
typical adsorption time of $\unit{3}{\milli \second}$ \cite{biance09}, which leads to mobile interfaces in most foam experiments. The fourth solution used a mixture of sodium lauryl-dioxyethylene sulfate (SLES), cocoamidopropyl betaine (CAPB), and myristic acid (MAc), specifically developed to produce rigid interfaces. Measures with the oscillating bubble method give $E_d < 1{\milli \newton \per \meter}$ for (T) solutions and   $E_d = \unit{320}{\milli \newton \per \meter}$ at \unit{0.2}{\hertz} for (D) \cite{biance09, golemanov08}.

\begin{table}
	 \caption{\label{tab:solutions}Solutions properties. Glycerol (10 vol. \%) is added to every solution, leading to a  viscosity $\eta=$ \unit{1.4}{\milli Pa\usk\second}. The TTAB critical micelle concentration ($c_{cmc}$) is $\unit{3.8}{\milli \mole \per \liter}$.}
 	\begin{ruledtabular}
 	\begin{tabular}{|c|c|c|}
 		solution & composition &  surface tension $\gamma$ \\ \hline
		T3 & TTAB  (\unit{14.8}{\milli \mole \per \liter} = 3 $c_{cmc}$) & \unit{35}{\milli \newton \per \meter}\\ 
		T13 & TTAB  (\unit{63.9}{\milli \mole \per \liter} = 13 $c_{cmc}$) & \unit{35}{\milli \newton \per \meter}  \\
		T25 & TTAB  (\unit{128}{\milli \mole \per \liter} = 25 $c_{cmc}$) & \unit{35}{\milli \newton \per \meter}  \\
		D & SLES  (0.30), CAPB (0.15)& \unit{23}{\milli \newton \per \meter}\\
& and MAc (0.05) (in wt. \%)& 
 	\end{tabular}
 	\end{ruledtabular}
 \end{table}

\begin{figure}[htp!]
\includegraphics[width=8.6cm]{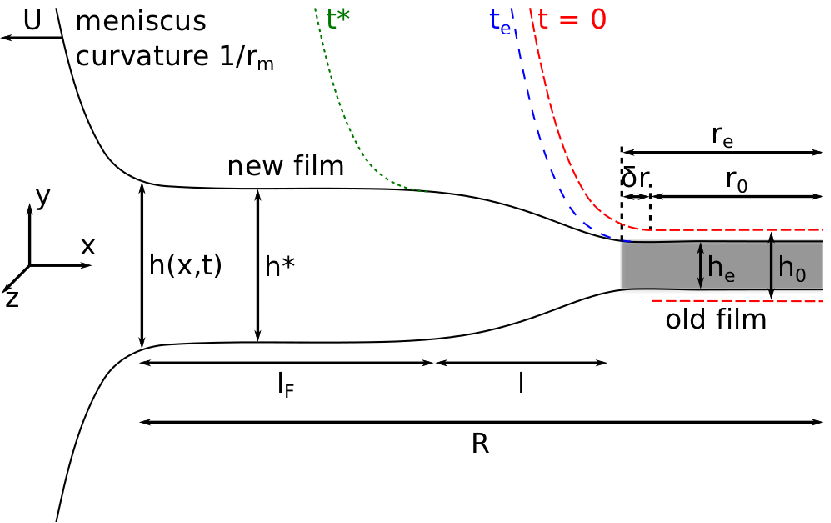}\\
\includegraphics[width=8.6cm]{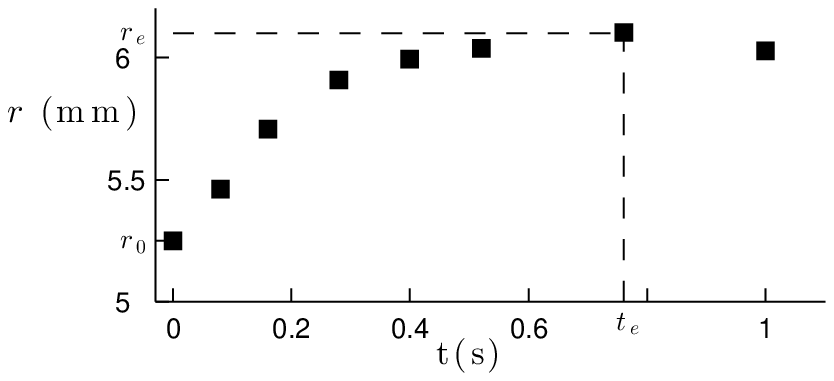}
\caption{Top: horizontal film profiles along a radius, from the meniscus to the film center, at different times. The dashed and dotted lines  are schematic representations of the film at $t=0$: initial time; $t=t_e$: end of the old film extension;  $t=t^*$: end of the new film transient. The solid line is the numerical result of Eq.~(\ref{eqn:QCons}) for a time $t> t^*$.
Bottom: radius of the old film $r$ as a function of time, for $U_s = \unit{5}{\milli \meter \per \second}$ and solution T13. It saturates at $r_e=r(t_e)$.
\label{fig:schTheo}}
\end{figure}

Upon retraction of the Plateau border, two successive regimes, schematized in Fig.~\ref{fig:schTheo}, are observed when using the TTAB solutions. During the first step ($0<t<t_e$), the film initially present (thereafter denoted ``old film'') undergoes a homogeneous extension, as attested by the motion of the thickness fluctuations.  
For $t>t_e$ (second step), the old film extension ceases and the subsequent increase in area is entirely compensated for by a thicker film extracted from the Plateau border, denoted ``new film''. 
 This new film consists of a transition region of characteristic length $l$, followed by a region of length $l_F$ and of approximately constant thickness, with a smooth maximum $h^*$.
As observed in Fig. \ref{fig:experiment} (bottom, right), these old and new films are well distinguishable during the entire experiment, and the film radius $R$ can be decomposed unambiguously  into three terms $R= r + l+ l_F$, with $r$ the old film radius.
For the solution (D), only the second step has been observed. 

 \begin{figure}[htp!]
	   		\includegraphics[width=8.6cm]{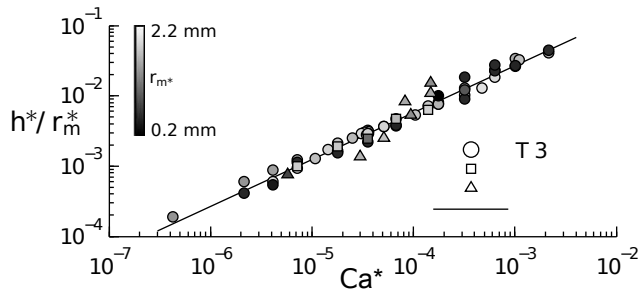}
			\caption{Thickness $h^*$ of the withdrawn film rescaled by the Plateau border radius as a function of the capillary number $\Ca^*$, for different surfactant solutions. The gray value of each symbol is proportional to $r_m^*$, from \unit{0.2}{\milli \meter} (black) to \unit{2.2}{\milli \meter} (white).  Data collapse on Frankel's law (solid line): $h_F/r_m^* = 2.68 \, \Ca^{*\,2/3}$.%
			\label{fig:frankel}}
		\end{figure} 

Let us first focus on the new film. 
We measured $h^*$ while varying systematically $U_s$ and $r_m$, for the different solutions. Figure~\ref{fig:frankel} shows that it strictly obeys  $h^*=h_F$, with $h_F= 2.68 \, r_m \, \Ca^{2/3}$ the thickness predicted by Frankel \cite{mysels}. Here, $\Ca = \eta U / \gamma$ is  the capillary number and $U$ the meniscus velocity. 
 Although Frankel's law has been observed previously when a soap film is withdrawn from a liquid bath, our measurements are to our knowledge the first made in a configuration relevant to foams, with Plateau borders as liquid reservoir. In particular, the radius of the Plateau border $r_m$ was varied over a decade.

The film of maximal thickness $h^*$ is extracted from the Plateau border at  time $t^*$. $Ca^*= \eta U(t^*)/\gamma$ is measured at this time.
The best accuracy ($\pm 10\%$) on  $r_m^*= r_m(t^*)$ was obtained by  fitting the meniscus profile at $t=0$ with its theoretical shape (taking into account corrections due to axisymmetry and  gravity) and by deducing $r_m^*$ from the meniscus volume conservation ($2 \pi R r^2_m $ is constant, neglecting the volume injected into the new film). Finally,  a fit of the form $h^* = K \, r_m^* \, \Ca^{*\,2/3}$ gives $K = 2.7 \pm 0.1$.

The transition region is not described by Frankel's theory, which only considers a steady state in the meniscus frame.
We thus developed an unsteady theoretical framework based on the same assumptions to predict $l$ \cite{cormier12}.  
The problem is solved in the Plateau border frame, assuming invariance in the $z$ direction. A thin film of initial uniform thickness $h_e(= h(t_e))$ is pulled out of a meniscus of radius $r_m$ at a velocity $U$ in the $x$-direction (see Fig.~\ref{fig:schTheo}, top). In the lubrication approximation, and assuming that the interfacial velocity is $U$ everywhere, the dimensionless equation of evolution written for the film half-thickness $H(X) = h(x)/h_F$ is  
\begin{equation}
		\label{eqn:QCons}
		\frac{\partial H}{\partial T} = -\frac{\partial}{\partial X} \left(H^3 \frac{\partial^3 H}{\partial X^3} +H  \right) \; ,
	\end{equation}
with $X=2 \,x(3 \Ca)^{1/3}/h_F$ and  $T= 2 \,t (3 \Ca)^{1/3}U/h_F$. 

Frankel's theory predicts the steady state shape of the film $H_F(X)$ by setting $\partial_t H_F = 0$ in (\ref{eqn:QCons}) and imposing $H_F(X \rightarrow + \infty) = 1$. On the meniscus side ($X \rightarrow - \infty$), the curvature of this solution reaches the  constant value 0.643. Matching with the meniscus curvature $1/r_m$ imposes  Frankel's law  $h_F= 2.68 r_m \Ca^{2/3}$.

\begin{figure}[htp!]
	   		\includegraphics[width=8.6cm]{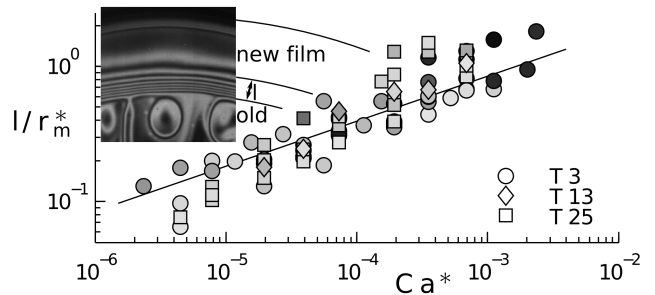}
			\caption{Length $l$ of the transition region between the initial film and Frankel's film (rescaled by $r_m^*$), as a function of $\Ca^*$ (color coding is identical to figure~\ref{fig:frankel}). The solid line is the numerical prediction $l/r_m^* = 8.9 \, \Ca^{*1/3}$. Inset: magnified view of the film featuring the interference pattern obtained with a monochromatic light. Each fringe increment  corresponds to a thickness variation of \unit{0.2}{\micro \meter}.
			\label{fig:profile}}
		\end{figure}

We solved the time-dependent equation using as initial condition a profile with the required curvature at $X \to - \infty$ (that is maintained constant in time) and a constant thickness $H_e <1$ at  $X \to + \infty$. A flat film of thickness $H = 1$ emerges from the meniscus and connects to the initial film through a well-defined transition region of length $l$ (defined as the length of the region encompassing 90\% of the height variation). 
The profile obtained for $H_e = 0.3$ and $T=20$ is plotted in Fig.~\ref{fig:schTheo}(top) for $X$ in the range [-8;22]: with these values of the parameters, $l =8.9  \, r_m \, \Ca^{1/3}$. 
Numerics showed that the variation of $l$ with $H_e$ (less than  20\% in the range $[0.2-0.8]$) and $T$ (less than 1\% in the range $[15-30]$) are well below our experimental dispersion,  hence we compared all  our experimental data with the previous expression, finding a good agreement (Fig.~\ref{fig:profile}). 

These results indicate that both the steady state thickness and the transition region of the new film exactly follow Frankel's theory. 
Our data thus evidence that all the tested solutions are able to generate incompressible interfaces. 
This incompressibility is ensured by surface tension gradients, mainly arising from variations in the surface excess of surfactants $\Gamma$, which balance bulk viscous stresses. 
For solution (D), these gradients form as soon as the interface stretches, independently of the subsurface concentration. In contrast, for (T) solutions, surface and bulk concentrations equilibrate faster than the experimental timescale.
Surface tension gradients thus imply  bulk  concentration gradients, which appear in our system because of confinement effects:  the film is thin enough that it does not contain much more surfactants than the interfaces, and is thus easily depleted \cite{quere98}. It is also large enough that diffusion from the Plateau border is negligible (the typical bulk diffusion time is $10^{5}$s).
The observed rigidification is thus a
striking illustration of the fact that surfactant behaviors do not
exclusively depend on their intrinsic properties, but also on the
length and time scales involved \cite{alvarez12,cantat12}.

The surface tension variation needed to pull the new film out of the Plateau border can be deduced from Frankel's calculation  \cite{mysels,cantat13}
	\begin{equation}
		\label{eqn:dGamma}
		\frac{	\delta \gamma }{\gamma} = \frac{\gamma_F - \gamma_P}{\gamma_P} = 3.8 \Ca^{2/3} \; , 
	\end{equation}
where $\gamma_P$ and $\gamma_F$ are the surface tensions in the Plateau border (where it is equal to the equilibrium surface tension of the foaming solution) and at the end of the new film ({\it i.e.} in the old film). The old film surface tension must thus increase to allow extraction of Frankel's film, with a relative variation in the range $[10^{-4}-10^{-2}]$ for  $\Ca$ in the range $[10^{-6}-10^{-3}]$.

The surface tension variation obeys $\delta \gamma =2 E_d \delta r/r_0$, with $\delta r=r_e-r_0$, as defined in Fig. \ref{fig:schTheo}. From Eq.~(\ref{eqn:dGamma}), we find $\delta r/r_0 \sim 10^{-2}$ at $\Ca=10^{-3}$ for (D) solutions, which is consistent with the absence of measurable extension of the old film.
In contrast, the $E_d$ value measured in unconfined geometry is irrelevant for (T) solutions. It would lead in the same conditions to   $\delta r/r_0 >1$, which is not compatible with the measured extension.

To predict the area increase of the old film for (T) solutions, we must take into account the absorption of surfactant from the volume on the newly created interface, which lowers the bulk concentration in the film and leads to an increase of its surface tension. Let us consider a film of initial radius $r_0$ and thickness $h_0$, with a bulk concentration of surfactant $c_0 = \alpha \, c_{cmc}$ ($\alpha > 1$). We consider surfactants with small adsorption time, for which bulk and surface concentrations remain at equilibrium, and we assume that there is neither solution nor surfactant exchange with the Plateau border during the entire stretching.
The surface tension is constant for $c>c_{cmc}$. 
Below the cmc,  $\gamma$ decreases with  $c$ in a way that is difficult to predict or measure with the required accuracy. A good agreement with the experimental data is obtained by assuming that  $\gamma = \gamma_{cmc} + K_\gamma (c - c_{cmc})^2$  for $c<c_{cmc}$, that is the simplest function smoothly matching the condition  $\gamma= \gamma_{cmc}$  for $c>c_{cmc}$.
The required surface tension variation is thus obtained for a concentration variation $\delta c/c_{cmc}=1- \alpha - \sqrt{\delta \gamma/K_\gamma}/c_{cmc}$.

A surfactant mass balance then relates  $\delta c$ to  $\delta r$. At first order in $\delta r$, $\delta r/r_0 = -h_0 \delta c/ (4 l_{cmc}\, c_{cmc})$, where $l_{cmc} = \Gamma_{cmc} /c_{cmc} \approx \unit{1}{\micro \meter}$. We used the fact that $\Gamma \approx \Gamma_{cmc} = \unit{3.6\, 10^{-6}}{\mole \per \meter \squared}$ for TTAB \cite{lu92}. Using Eq.~(\ref{eqn:dGamma}) we obtain
	\begin{equation}
		\label{eqn:dr}
		\frac{\delta r}{r_0} = \frac{h_0}{4 l_{cmc}} \left(\alpha - 1 + \sqrt{\frac{3.8 \gamma_{cmc}}{K_\gamma c_{cmc}^2}} \Ca^{1/3}\right) \; . 
  \end{equation}

\begin{figure}[htp!]
			\includegraphics[width=8.6cm]{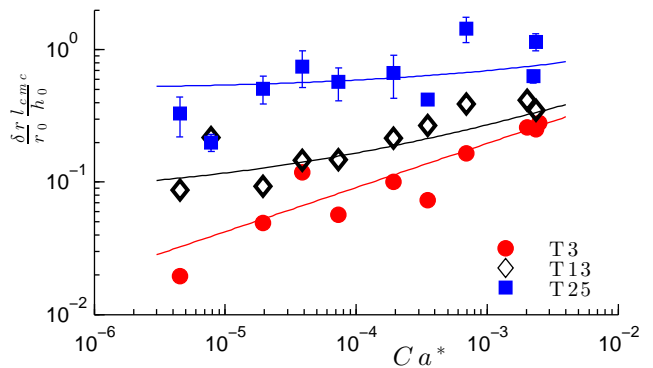}
			\caption{Old film maximal extension $\delta r= r_e-r_0$ rescaled by $r_0 h_0/l_{cmc}$  as a function of $\Ca$ for the three TTAB solutions (averaged over several 3 series of various $r_m$ for the T25, the error bars representing the standard deviation).  Solid lines are the best fits of each curve with the prediction (\ref{eqn:dr}) using $K_\gamma = \unit{0.15}{\milli \newton \usk \meter^5 \per \mole^2} $, $\alpha_3 = 1$, $\alpha_{13} = 1.3$ and $\alpha_{25} = 3.5$.
			\label{fig:extension}}
		\end{figure}

We plotted on Fig.~\ref{fig:extension} the old film extensions measured as a function of $\Ca$, for the three solutions of type (T). The parameter $\alpha$ of eq. \ref{eqn:dr} was fitted for each solution and a good agreement was obtained for $\alpha_3 = 1$, $\alpha_{13} = 1.3$ and $\alpha_{25} = 3.5$. These values are lower than the initial bulk concentrations ($\alpha = 3$, 13 and 25 respectively), which can be explained by the initial drainage process, known to reduce the concentration in the film \cite{sonin93,breward02}. The parameter $K_\gamma$ was optimized globally for the three different solutions, with a best fit value $K_\gamma = \unit{0.15}{\milli \newton \usk \meter^5 \per \mole^2}$. 

In summary, we evidence in this Letter a homogeneous extension regime of surfactant films, at small strain. This is in agreement with the linear relationship between stress and strain rate measured on surfactant films subject to small amplitude oscillatory forcing \cite{besson07}. At larger strain, we evidence a second regime, which is perfectly described by a rigid interface model. This is the Frankel's regime, known to lead to a sub-linear relation between stress and strain rate \cite{schwartz87}. 
The elementary deformation that we study may contribute significantly to the global dissipation in foam under steady shear, especially at high shear rate, when the dissipation directly associated to topological rearrangements is not dominant. Our results are therefore an important step  towards  improving models of foam rheology. 

\begin{acknowledgments}
We thank A.-L. Biance, E. Lorenceau and A. Saint-Jalmes for fruitful discussions and  J.-C. Potier for technical support. J. S. acknowledges financial support from R\'egion Bretagne (CREATE MOUSPORE), Agence Nationale de la Recherche (ANR- 11-JS09-012-WOLF), and University Rennes 1.  
\end{acknowledgments}

%\bibliographystyle{/home/cantat/isabelle/bib/physrev}
%\bibliography{/home/cantat/isabelle/bib/bib}

\begin{thebibliography}{0}%
\makeatletter
\providecommand \@ifxundefined [1]{%
 \@ifx{#1\undefined}
}%
\providecommand \@ifnum [1]{%
 \ifnum #1\expandafter \@firstoftwo
 \else \expandafter \@secondoftwo
 \fi
}%
\providecommand \@ifx [1]{%
 \ifx #1\expandafter \@firstoftwo
 \else \expandafter \@secondoftwo
 \fi
}%
\providecommand \natexlab [1]{#1}%
\providecommand \enquote  [1]{``#1''}%
\providecommand \bibnamefont  [1]{#1}%
\providecommand \bibfnamefont [1]{#1}%
\providecommand \citenamefont [1]{#1}%
\providecommand \href@noop [0]{\@secondoftwo}%
\providecommand \href [0]{\begingroup \@sanitize@url \@href}%
\providecommand \@href[1]{\@@startlink{#1}\@@href}%
\providecommand \@@href[1]{\endgroup#1\@@endlink}%
\providecommand \@sanitize@url [0]{\catcode `\\12\catcode `\$12\catcode
  `\&12\catcode `\#12\catcode `\^12\catcode `\_12\catcode `\%12\relax}%
\providecommand \@@startlink[1]{}%
\providecommand \@@endlink[0]{}%
\providecommand \url  [0]{\begingroup\@sanitize@url \@url }%
\providecommand \@url [1]{\endgroup\@href {#1}{\urlprefix }}%
\providecommand \urlprefix  [0]{URL }%
\providecommand \Eprint [0]{\href }%
\providecommand \doibase [0]{http://dx.doi.org/}%
\providecommand \selectlanguage [0]{\@gobble}%
\providecommand \bibinfo  [0]{\@secondoftwo}%
\providecommand \bibfield  [0]{\@secondoftwo}%
\providecommand \translation [1]{[#1]}%
\providecommand \BibitemOpen [0]{}%
\providecommand \bibitemStop [0]{}%
\providecommand \bibitemNoStop [0]{.\EOS\space}%
\providecommand \EOS [0]{\spacefactor3000\relax}%
\providecommand \BibitemShut  [1]{\csname bibitem#1\endcsname}%
\let\auto@bib@innerbib\@empty
%</preamble>
\end{thebibliography}%


\begin{thebibliography}{10}

\bibitem{britan13}
A.~Britan {\em et~al.},
\newblock Shock Waves {\bf 23}, 5 (2013).

\bibitem{delprete13}
E.~Del~Prete, A.~Chinnayya, L.~Domergue, A.~Hadjadj, and J.-F. Haas,
\newblock Shock Waves {\bf 23}, 39 (2013).

\bibitem{stevenson}
P.~Stevenson, editor,
\newblock {\em Foam Engineering. Fundamentals and Applications} (Wiley, 2012).

\bibitem{cohenaddad13}
S.~Cohen-Addad, R.~H\"ohler, and O.~Pitois,
\newblock Annu. Rev. Fluid. Mech. {\bf 45}, 241 (2013).

\bibitem{kraynik87}
A.~M. Kraynik and M.~G. Hansen,
\newblock J. Rheol. {\bf 31}, 175 (1987).

\bibitem{schwartz87}
L.~W. Schwartz and H.~M. Princen,
\newblock J. Colloid Interface Sci. {\bf 118}, 201 (1987).

\bibitem{buzza95}
D.~Buzza, C.-Y. Lu, and M.~E. Cates,
\newblock J. Phys. II (France) {\bf 5}, 37 (1995).

\bibitem{denkov08}
N.~D. Denkov, S.~Tcholakova, K.~Golemanov, K.~P. Ananthapadmanabhan, and
  A.~Lips,
\newblock Phys. Rev. Lett. {\bf 100}, 138301 (2008).

\bibitem{durand06}
M.~Durand and H.~A.~Stone,
\newblock Phys. Rev. Lett. {\bf 97}, 226101 (2006).

\bibitem{denkov06}
N.~D. Denkov, S.~Tcholakova, K.~Golemanov, V.~Subramanian, and A.~Lips,
\newblock Colloids Surf. A {\bf 282}, 329 (2006).

\bibitem{besson07}
S.~Besson and G.~Debr\'egeas,
\newblock Eur. Phys. J. E {\bf 24}, 109 (2007).

\bibitem{biance09}
A.~L. Biance, S.~Cohen-Addad, and R.~H\"ohler,
\newblock Soft Mat. {\bf 5}, 4672 (2009).

\bibitem{dollet10}
B.~Dollet and I.~Cantat,
\newblock J. Fluid Mech. {\bf 652}, 529 (2010).

\bibitem{biance11}
A.-L. Biance, A.~Delbos, and O.~Pitois,
\newblock Phys. Rev. Lett. {\bf 106}, 068301 (2011).

\bibitem{cantat12}
I.~Cantat and B.~Dollet,
\newblock Soft Mat. {\bf 8}, 7790 (2012).

\bibitem{denkov05}
N.~D. Denkov, V.~Subramanian, D.~Gurovich, and A.~Lips,
\newblock Colloids Surf. A {\bf 263}, 129 (2005).

\bibitem{marze08}
S.~Marze, D.~Langevin, and A.~Saint-Jalmes,
\newblock J. Rheol. {\bf 52}, 1091 (2008).

\bibitem{denkov09b}
N.~D. Denkov, S.~Tcholakova, K.~Golemanov, K.~P. Ananthpadmanabhan, and
  A.~Lips,
\newblock Soft Mat. {\bf 5}, 3389 (2009).

\bibitem{golemanov08}
K.~Golemanov, S.~Tcholakova, N.~D. Denkov, M.~Vethamuthu, and A.~Lips,
\newblock Langmuir {\bf 24}, 9956 (2008).

\bibitem{besson08}
S.~Besson, G.~Debr\'egeas, S.~Cohen-Addad, and R.~H\"ohler,
\newblock Phys. Rev. Lett. {\bf 101}, 214504 (2008).

\bibitem{mysels}
K.~J. Mysels, K.~Shinoda, and S.~Frankel,
\newblock {\em Soap films: Study of their thinning and a bibliography}
  (Pergamon, New-York, 1959).

\bibitem{geminard04}
J.-C. G\'eminard, A.~\.{Z}ywoci\'nski, F.~Caillier, and P.~Oswald,
\newblock Phil. Mag. Lett. {\bf 84}, 199 (2004).

\bibitem{chomaz01}
J.~M. Chomaz,
\newblock J. Fluid Mech. {\bf 442}, 387 (2001).

\bibitem{cormier12}
S.~L. Cormier, J.~D. McGraw, T.~Salez, E.~Rapha\"el, and K.~Dalnoki-Veress,
\newblock Phys. Rev. Lett. {\bf 109}, 154501 (2012).

\bibitem{alvarez12}
N.~J. Alvarez, D.~R. Vogus, L.~M. Walker, and S.~L. Anna,
\newblock J. Colloid Interface Sci. {\bf 372}, 183 (2012).

\bibitem{quere98}
D.~Qu\'er\'e and A.~de\ Ryck,
\newblock Ann. Phys. {\bf 23}, 1 (1998).

\bibitem{cantat13}
I.~Cantat,
\newblock Phys. Fluids {\bf 25}, 031303 (2013).

\bibitem{lu92}
J.~R. Lu {\em et~al.},
\newblock J. Phys. Chem. {\bf 96}, 10971 (1992).

\bibitem{sonin93}
A.~A. Sonin, A.~Bonfillon, and D.~Langevin,
\newblock Phys. Rev. Lett. {\bf 71}, 2342 (1993).

\bibitem{breward02}
C.~J.~W. Breward and P.~D. Howell,
\newblock J. Fluid Mech. {\bf 458}, 379 (2002).

\end{thebibliography}

\end{document}